# Investigation of the aptness of newly developed epoxy-based equivalent tissues for newborn and 5-years old in paediatric radiology

Nabeel Ibrahim Ashour[1,2], Muhammad Fahmi Rizal Abdul Hadi[1], Nurul Ab. Aziz Hashikin[1,*], Mohammed Ali Dheyab[1], Ahmed Sadeq Musa[1,3], Nik Noor Ashikin Nik Ab Razak[1] and Mohd Zahri Abdul Aziz[4]

[1] School of Physics, Universiti Sains Malaysia, 11800, Penang, Malaysia

[2] Department of Physics, College of Science, University of Kerbala, 56001 Kerbala, Iraq

[3] Department of Physiology and Medical Physics, College of Medicine, University of Kerbala, 56001 Kerbala, Iraq

[4] Department of Biomedical Imaging, Advanced Medical & Dental Institute, Universiti Sains Malaysia, 13200, Penang, Malaysia

*  **Corresponding author**: hashikin@usm.my

**Abstract**

The varied radiological applications of tissue equivalent (TE) materials encompass quality checks, diagnostic imaging and dose evaluations. Nevertheless, the availability of compounds representative of paediatric patient tissues for scientific use in lower diagnostic photon energy spectra is limited. In this study, several TE substitutes were developed which replicate the radiographic characteristics of human tissue within these energy ranges, i.e. TE materials for neonatal soft tissue (ESST-NB), neonatal skeletal tissue (ESTB-NB), and the equivalent tissue types representative of a 5 year old child (ESST and ESBT, respectively). The ORNL stylised computational model series was used as a source for the desired elemental proportions. The density, effective atomic number, CT numbers and electron densities calculated for the developed tissue substitutes approximated those of the phantom system used as a reference. Additionally, in keeping with the material choice and production limitations, as close correlations as possible were achieved for all the materials in relation to the reference data for mass densities, mass attenuation coefficients and mass energy-absorption coefficients. The TE substitutes for the newborn over an energy range of 47 keV to 66 keV exhibited maximum discrepancies for $\mu/\rho$ of +1.6% to –3.01%, and for $\mu_{en}/\rho$ of +1.15% to –1.4% in relation to the ORNL reference samples. The respective equivalent data ranges were +1.09 % to – 3.02% and +1.92% to –2.53% for the TE materials representative of a 5-year-old. Given the excellent concordance achieved between the newly constructed TE materials and the reference data, these compounds can subsequently be utilised to create physical phantoms representative of tissue types in neonates and children aged 5 years.

# 1. INTRODUCTION

For years, in medicine, ionizing radiation is used in paediatric radiology for diagnostic purposes such as imaging of fetuses, babies, children, and adolescents. Due to their highly proliferating cells and longer life expectancy for children, therefore, must be kept them away from ionizing radiation effects more than adults [1, 2]. Hence, it is important to evaluate the radiation protection of children and adolescents different from adults, as their tissues' radiation absorption properties are not the same. One of the tool that can be used to measure the radiation dose is by using a phantom which is a human tissue equivalent material.

Phantoms are essential tools for representing the patient and are used in radiology physics to determine the optimum radiation dose or image quality tests [3]. These phantoms should be identical to the tissues of the patient almost. Therefore, the phantom design requires the careful selection of tissue substitute materials. The materials must closely match the volume, density, and chemical composition characteristics of the represented human tissue for a proper radiological response at the energy of interest. The analysis of the dose in pediatric radiology is of interest since growing tissue is generally considered more sensitive to radiation and in most examinations, a larger portion of the child's body is included in the primary beam [4]. Age-appropriate phantoms are required for accurate analysis of pediatric dosimetry. An adult phantom or even a scaled-down version of an adult phantom is not appropriate for studying pediatric examinations or for comparing the results to those from mathematical modeling [5].

In 1906, Kienboch suggested that for research purposes, water could be considered to be a comparable physical entity to muscle [6]. Water is present in a liquid phase, so it is not a consistently suitable substance for application to studies relating to medical physics. Since this initial concept, scientists have continued to look for substances that are representative of human tissues. There are a number of easily accessible substances, including acrylic and aluminium, which can be employed for radiology diagnostic quality assurance. However, these are not often ideal for use in dosimetry studies, particularly those relevant to the paediatric radiology domain in which energies of <120 keV are employed.

In the last century, a number of advances and improvements were carried out with respect to the development of tissue equivalent (TE) materials. Currently, there are four varieties of phantom materials in use which are categorised in accordance with the base material. White et al. and others mainly developed an epoxy-resin-based material [3, 7-17], Hermann et al. used a polyethylene-based method [18], Homolka's group made phantoms from fine polymer powders, e.g. polyethylene, polypropylene, polystyrene or polyurethane [19], and Suess, Iwashita and others used polyurethane resin [20, 21]. Several issues have been noted in previous studies. These comprise: (i) most of the TE materials mentioned in previous studies have been designed to match the tissue attenuation characteristics at relatively high energies (around 1 MeV and higher), while there are few studies concerned with constructing paediatric TE to match the attenuation characteristics at low diagnostic energies < 120 keV; (ii) the basis for tissue equivalent materials for many years has been mixtures based on common unmodified epoxy resins. e.g. Araldite GY-6010 [13]. The epoxy resin may be combined with a variety of other molecular compounds to achieve the desired mass attenuation coefficient and density, but it was some of this type of epoxy resin had been halted manufacture from the chemical company [22]; (iii) the cost prohibitively expensive for various commercial paediatric phantoms that which used as a stander reference in various medical application [23-25]

Therefore, as a part of our efforts to construct paediatric TE substitutes, the present study selects TE materials based on a new epoxy resin, and all the components were designed in certain proportions to be equivalent from a radiographical perspective to the soft tissue and skeletal components of a neonatal and 5 years old subject. The criteria described by Cristy and Eckerman for the Oak Ridge National Laboratory (ORNL) stylised model series [13, 26] were taken as reference targets for the composition of the tissues used in this study.

It is anticipated that the de novo tissue substitutions designed at University Sains Malaysia (USM) will exhibit a precise response at lower energies. Comparisons of these TE substances are presented in the current study against a range of alternative TE substances in contemporary use for the acquisition of quality assurance parameters and for research relating to clinical dosimetry. The ultimate goal of this research is to design 3-dimensional physical tomographic phantoms representative of tissues from neonatal and 5-year old paediatric subjects in accordance with the basic properties stated in Publication 143 of the International Commission on Radiological Protection [27].

## 2. MATERIALS AND METHODS

*2.1 Phantom Materials for newborn and toddler*

All the TE substances referred to were produced using an epoxy resin base for which the mass density was modified with phenolic microspheres. The epoxy resin was procured in liquid phase and purchased from (Huntsman Corporation; USA). Use of epoxy resin in combination with a number of fillers that have been purchased from Sigma (Sigma Aldrich; Merck, Missouri, USA). Since the company (Huntsman Corporation; USA) halted manufacture synthesising the epoxy resin, Araldite GY 6010. Thus, there was a pressing requirement to source a further epoxy resin option. In place of Araldite GY6010, used by previous researchers [3, 13], Araldite GY250 was suggested for the current study. This compound has several distinctive characteristics and the selected filler substances are accessible, reasonably straightforward to produce, and have significant longevity [5]. The constituents for the individual tissue substitutes were modified as indicated in Table 1.

*2.1.1. Equivalent substitute-soft tissue for the newborn*

The parameters described by Cristy and Eckerman [26] were used as a benchmark for the generation of radiographically equivalent substitute-soft tissue for the newborn (ESST–NB). The latter was produced from an Araldite GY-250 base as an epoxy resin, in combination with the hardening agent, Jeffamine T-403. The ratios of these components were approximate to those utilised by White et al. [11] and Jones et al. [13]. However, for the purposes of the present study, these were modified and enhanced by substituting the originally used Araldite GY-6010 with Araldite GY-250. The two compounds exhibited similar characteristics, although the viscosity of Araldite GY-250 was lower than that for the Araldite GY-6010. Polyethylene, silicon dioxide, and magnesium oxide were employed as filler agents, and the required mass density was achieved with the inclusion of phenolic microspheres.

*2.1.2. Equivalent substitute-bone tissue for the newborn*

The substances innovated at USM to be equivalent to skeletal tissue are designed to reflect a uniform combination of skeletal trabeculae and bone marrow, i.e. cortical and trabecula spongiosa, respectively. The parameters published by Cristy and Eckerman [26] were also utilised as a benchmark for the production of radiographically equivalent substitute-bone tissue for the newborn (ESBT–NB). In a similar manner to the ESST-NB, ESBT-NB was synthesised from an Araldite GY-250 base together with the hardening agent, Jeffamine T-403. These were admixed in a ratio which approximated that originally described [11, 13]. Polyethylene, silicon dioxide, magnesium oxide and calcium carbonate were utilised as filler agents in the ratios necessary to create corresponding values to the ORNL standard for neonatal bone tissue within the required diagnostic energy spectrum for mass density, mass attenuation coefficients and mass energy absorption coefficients.

### 2.1.3. Equivalent substitutes-soft tissue and bone tissue for 5-year-old subject

The elemental constituents with respect to soft tissue and uniform skeletal tissue described in the ORNL model series vary according to subject age when more mature than the neonate. Values are given for subjects aged 1, 5, 10 and 15 years, and for adulthood. Thus, a distinctive elemental composition is only defined for the newborn model. We added zinc oxide as a new filler and the modifications to the ratios of the filling agents for the manufacture of ESST–NB and ESBT–NB were applied in order to produce empirical TE replacements for application to phantom generation for the representation of older subjects. The acronyms, ESST and ESBT, in the absence of the newborn (NB) suffix, are used to indicate the TE material for this population.

**Table 1.** Material compositions of the USM tissue- equivalent substitutes.

| Material | Material composition (% by mass) | | | |
| --- | --- | --- | --- | --- |
| | ESST-NB | ESBT-NB | ESST | ESBT |
| Araldite GY-250 | 50.70 | 46.60 | 49.04 | 40.20 |
| Jeffamine T-403 | 20.30 | 19.90 | 20.94 | 15.30 |
| Polyethylene | 8.70 | | 9.24 | |
| Magnesium oxide | 15.50 | | 16.24 | |
| Silicon dioxide | 1.30 | 15.00 | 0.54 | 22.50 |
| Phenolic microspheres | 3.50 | | 2.5 | |
| Zinc oxide | | | 1.5 | 1.8 |
| Calcium carbonate | | 17.50 | | 20.20 |
| Polyvinyl chloride | | 1.00 | | |

### 2.2 Preparation of Phantom's Materials

The TE materials were synthesised using a technique resembling the process described by White at al. [11]. Following weighing, the components were combined in a pre-defined order so as to ensure that they were thoroughly admixed. The dry constituents were added to the correct amount of Araldite GY-250; these were followed by phenolic microspheres when required and, ultimately, by the hardening agent. The admixture was then agitated by hand in order to ensure that all the components were fully mixed to form a dough consistency. Mechanical stirring, utilising an electromechanical stirrer which included a paint agitator, was only commenced at this juncture so as to avoid losing any of the dry constituents. The stirring duration for the individual soft tissue mixtures was between 3 and 5 minutes. Any trapped air was extracted by subjecting the mixture to a vacuum system for 5 minutes on each occasion. The vacuum process was not applied to the skeletal formulations; these had a lower viscosity which facilitated the liberation of any collections of air. The next stage was the addition of surfactants and foaming agents which were blended into the material using the mechanical stirrer. Following decantation into release-treated moulds, the admixture was then left alone to foam until curing was achieved. An in-depth illustration of the synthesis technique is presented in Figure 1.

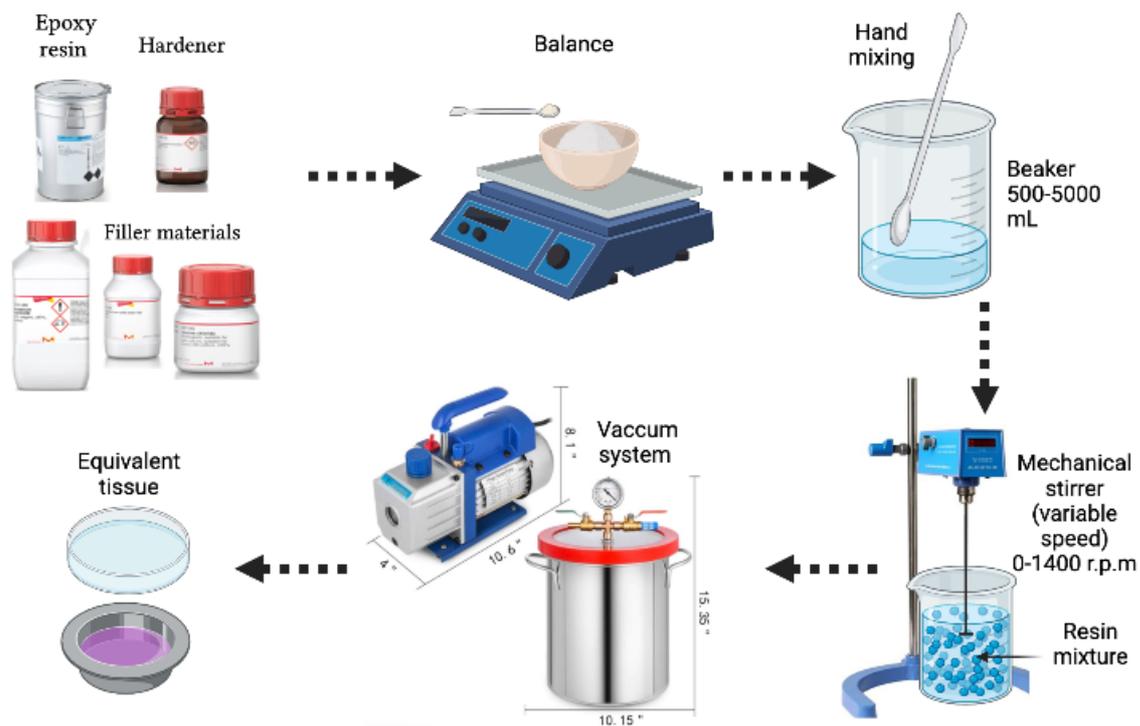

**Figure 1.** Methodology and equipment for manufacturing equivalent substitutes of the soft and bone tissues.

## *2.3 Radiological Properties of tissue equivalent material*

### *2.3.1 Mass density*

In order to attain the precise density of the TE substitutes, several experiments were run utilising various proportions of Araldite GY-250 and filler agents. During the curing and admixing of the material there are a several possible chemical modifications which may arise, and so obtaining a straightforward weighted average of the constituent densities is not possible. An autonomous measurement of the mass densities for the individual de novo TE substitutes was therefore necessary before they could be used in the phantom build. This quality check was carried out for any TE substitute which had a density greater than that of water, i.e. $\rho = 0.9975 \text{ to } 0.9980 \text{ } gm \text{ } cm^{-1}$, depending on temperature. This was based on the Archimedean principle:

$$\rho_{material} = \frac{M}{B} \times \rho_{water},  \qquad (1)$$

where m and B represent the measured material's dry mass and buoyancy in water, respectively. This is expressed as the difference in the material's dry mass and the displaced water mass following submergence of the material.

*2.3.2 Effective atomic number, Zeff*

The configuration of the TE materials was utilised to identify the elemental components using EDX analysis. The latter facilitated the energy beam to be used to produce high quality images with a 40,000-fold level of magnification. The test samples were fashioned into 0.5 cm-sided cubes; conductivity was promoted by adding a platinum veneer, which diminished thermal injury, enhanced the secondary signal and circumvented the acquisition of disrupted image quality. The specimen was secured onto a holder and then subjected to scanning electron microscopy which was performed in a vacuum. The proportions of the various comprising elements were acquired, with the exception of hydrogen atoms, utilising EDX; CHN/O analysis provided information relating to the latter. Founded on the Dumas technique, CHN/O analysis provided data on the carbon (C), hydrogen (H), nitrogen (N) and oxygen (O) constituents of the test sample. Equation (2) facilitated computation of the effective atomic number, $Z_{eff}$:

$$Z_{eff} = \left[ \sum_{i=1}^{n} (w_i z_i^m) \right]^{\left(\frac{1}{m}\right)} \qquad (2)$$

where $w_i$ and $z_i$ denote the electron fraction and atomic number, respectively, for the sample's $i^{th}$ element, m is the experimental coefficient for biological materials, and $Z_{eff}^{PE}$ is described by a more representative value for photoelectric absorption [28, 29] as defined in Equation (3):

$$Z_{eff}^{PE} = \sqrt[3.5]{\sum a_i z_i^{3.5}} \qquad (3)$$

and by $a_i = \left[ \left( \frac{w_i z_i}{A_i} / \sum_i \left( \frac{w_i z_i}{A_i} \right) \right) \right]$, where $A_i$ represents the mass number of element $i$.

*2.3.3 CT number and electron density (ED)*

This process was performed at the Advanced Medical and Dental Institute, university Sains Malaysia. A tube current of 150 mA in combination with a diagnostic energy of 120kVp was applied for the scanning of CIRS phantoms with the TE specimens and plug phantoms. The quality of the images was optimised, and the radiation dose minimised by the application of Adaptive Iterative Dose Reduction 3D (AIDR 3D). A filter kernel (FC18) in the absence of extra beam hardening correction was used for image reconstruction with a slice thickness of 0.75 mm. Software integral to the computerised tomography (CT) system (SOMATOM Definition AS, Siemens, Germany) was employed in order to establish the sample density distribution which is tightly linked with the CT number. The density plugs phantom numbers facilitated the plotting of the ED curve; the CIRS phantom provided the ED values. Equation 4 illustrates calculation of the ED value:

$$ED = 0.001657\, CT + 3.369 \qquad (4)$$

where the test sample's CT number is indicated by CT.

### *2.3.4 Linear attenuation coefficient based on CT number*

A CT scanner (SOMATOM Definition.AS, Siemens, Germany) was used to image the manufactured specimens and the CIRS phantoms for this research. Average characteristics of attenuation and the chemical components of the region scanned can be implied by the CT number. Thus, it is possible to differentiate between two distinct materials from the variation between the values of the latter [30]. A spiral scan mode was utilised for cross-sectional CT image acquisition, with tube voltages of 70, 75, 80, 85, 90, 95, 100 and 120 kVp which correlated with the respective kiloelectron average energy values of 47, 49, 52, 54, 56, 58, 60, and 66 keV [31]. Automatic tube current modulation (mAs) was applied. Circular regions of interest (ROI) were established, for which CT number data were generated. Subsequently, computation of the linear attenuation coefficient was carried out using the empirical equation for the CT number which is expressed as:

$$\mu_{tissue} = \left(CT\ number \times \frac{\mu_w}{1000}\right) + \mu_w \qquad (5)$$

where $\mu_{tissue}$ indicates the tissue linear attenuation coefficient in a pixel for which the CT number is measured and $\mu_w$ represents a predetermined linear attenuation coefficient of water [32].

### *2.3.5 Mass attenuataion coefficient*

The mass attenuation coefficients of USM TE substitutes were calculated based on equation 6 by using the linear attenuation value from the section 2.3.4 and density from section 2.3.1. The mass attenuation coefficients values were then compared to the reference values of ORNL reference by dividing the equation 6 with 7. The mass energy absorption coefficients were calculated based on equation 8 for each TE substitute elemental value and then compared to the respected reference values of ORNL

$$\text{Mass attenuation coefficients} = \mu/\rho \tag{6}$$

$$\left(\frac{\mu}{\rho}\right)_{ORNL\ tissue} = \sum_i \omega_i \left(\frac{\mu}{\rho}\right)_i \tag{7}$$

$$\left(\frac{\mu_{en}}{\rho}\right)_{ORNL,\ TE\ Material} = \sum_i \omega_i \left(\frac{\mu_{en}}{\rho}\right)_i \tag{8}$$

where $\omega_i$ is the mass fraction of element $i$ in the TE material or ORNL reference sample.

Elemental reference values for μ/ρ and μ$_{en}$/ρ were extracted from previous publications [32, 33] for ORNL calculation (equation 7 and 8). In accordance with Attix [28], the more appropriate expression of the weighting factors referred to in Eq. (6) is $(1 - g_i)\,\omega_i$, where $g_i$ reflects the radiation yield fraction for element $i$ in the TE substance. However, the $g_i$ values are practically zero in the photon energy spectrum relevant to this research, i.e. <120 kVp, for all the included elements.

### *2.4 Comparison of USM TE substitutes to other reference materials*

The radiation interaction coefficients of TE substitutes (ESST-NB, ESBT-NB, ESST, ESBT) were compared to the acrylic, as well as aluminium, MS11, IB1 and SB5. Proposed by White et al. [11], aluminium, MS11, IB1 and SB5 comprise replacement substances based on epoxy resin and so they were felt to be ideal materials against which to compare those prepared at USM. MS11, IB1 and SB5 were designed to correspond to muscular tissue, a typical combination of osseous bone trabeculae and red marrow reflecting the cancellous skeletal interior spongiosa, i.e. an osseous to soft tissue ratio of 22.4% to 77.6%, and material which represented cortical bone, respectively [11]. Since polystyrene, a soft tissue substitute, is rarely employed for clinical radiological diagnosis apart from its application for scatter measurements, it was not included in this study [34, 35]. Additionally, computations indicated that its equivalence to the soft tissue in the ORNL reference samples was less than for acrylic within the relevant energy spectrum, i.e. 47 keV-66 keV. Copper, which has been used to match skeletal tissue, was also discarded from the comparative study as aluminium provides a closer correspondence with the ORNL bone reference sample over the desired energy range. Thus, the final comparison included acrylic and aluminium as firstly, they are frequently employed for quality checks in relation to diagnostic apparatus and can be incorporated within subject representative phantoms [36, 37], secondly, they are common components used in phantoms for both computed and digital radiography [37-40], and finally, for CT imagers, acrylic is the reference substance utilised for the confirmation of quality parameters [41, 42].

# 3  Results

## *3.1 Density and effective atomic number*

The ultimate constituents comprising the four substances utilised in this study are shown in Table 1. A comparison of the component elements and mass densities for the TE materials for the newborn patient, i.e. ESST-NB, ESBT-NB, and the matching ORNL reference specimens is presented in Table 2 [26]. In the lower portion of this table, the effective atomic number is indicated in two formats, i.e. $Z_{eff}$ and $Z_{eff}^{PE}$. These represent a mass-weighted average of the elemental atomic numbers and a pre-determined parameter which is more reflective of photo electric absorption described by Attix [28] and Johns and Cunningham [29], respectively. the $Z_{eff}$ values for ESST-NB and ESBT-NB show discrepancies with respect to the reference value of 2% and 3%, respectively, and $Z_{eff}^{PE}$ values for the same tissues show discrepancies of 1% and 0.6%, respectively. In the course of generating the TE materials, these two values were recognised as excellent indicators of the degree of mass attenuation coefficient and mass energy absorption-coefficient correspondence with the ORNL reference samples.

Table 2. Elemental composition and effective atomic numbers for the USM newborn equivalent tissue and the corresponding reference tissue compositions for the ORNL newborn.

| Element | Elemental composition (% by mass) | | | |
|---|---|---|---|---|
| | ESST-NB | Reference soft tissue | ESBT-NB | Reference bone tissue |
| H | 8.26 | 10.625 | 6 | 7.995 |
| C | 40.72 | 14.964 | 41.73 | 9.708 |
| N | 3.77 | 1.681 | 1.5 | 2.712 |
| O | 40.97 | 71.830 | 34.09 | 66.811 |
| Na | 0.15 | 0.075 | | 0.314 |
| Mg | 4.69 | 0.019 | | 0.143 |
| Si | 0.97 | | 8.23 | |
| P | | 0.179 | | 3.712 |
| S | | 0.240 | | 0.314 |
| Cl | 0.27 | 0.079 | 0.22 | 0.140 |
| K | 0.12 | 0.301 | | 0.148 |
| Ca | 0.08 | 0.003 | 8.23 | 7.995 |
| Fe | | 0.004 | | 0.008 |
| Density (g/cm³) | 1.04 | 1.04 | 1.22 | 1.22 |
| $Z_{eff}$ | 6.86 | 7.02 | 8.23 | 8.5 |
| $Z_{eff}^{PE}$ | 7.65 | 7.55 | 10.91 | 10.84 |

The equivalent values of elemental composition and mass density for the dosimetry phantom TE materials representing a 5-year-old subject are listed in Table 3. The reference values are extracted from the ORNL model series, but with the exception of data for the neonate. The mass densities of interest for the four TE materials were in close proximity. the $Z_{eff}$ values for ESST and ESBT show discrepancies with respect to the reference value of 3% and 2%, respectively, and $Z_{eff}^{PE}$ values for the same tissues show discrepancies of 1% and 2%, respectively.

**Table 3.** Elemental composition and effective atomic numbers for the USM equivalent tissues of 5-year-old phantom and their corresponding reference tissue compositions of ORNL model at similar age.

| Element | Material composition (% by mass) | | | |
|---|---|---|---|---|
| | ESST | Reference soft tissue | ESBT | Reference bone tissue |
| H | 7.97 | 10.454 | 5.11 | 7.337 |
| C | 54.62 | 22.663 | 42.81 | 25.475 |
| N | 1.54 | 2.490 | 1.52 | 3.057 |
| O | 30.09 | 63.525 | 33.92 | 47.893 |
| F | | | | 0.025 |
| Na | 0.07 | 0.112 | | 0.326 |
| Mg | 4.99 | 0.013 | | 0.112 |
| Si | 0.45 | 0.030 | 7.148 | 0.002 |
| P | | 0.134 | | 5.095 |
| S | | 0.204 | | 0.173 |
| Cl | 0.1 | 0.133 | 1.1 | 0.143 |
| K | | 0.208 | | 0.153 |
| Ca | 0.16 | 0.024 | 8.37 | 10.190 |
| Fe | | 0.005 | | 0.008 |
| Zn | 0.01 | 0.003 | 0.022 | 0.005 |
| Rb | | 0.001 | | 0.002 |
| Zr | | 0.001 | | |
| Sr | | | | 0.003 |
| Pb | | | | 0.001 |
| Density (g/cm³) | 1.04 | 1.04 | 1.44 | 1.4 |
| $Z_{eff}$ | 6.59 | 6.86 | 8.30 | 8.5 |
| $Z_{eff}^{PE}$ | 7.35 | 7.43 | 11.02 | 11.36 |

*3.2 CT number and electron density (ED)*

A marked alteration in the CT number was observed in relation to the high-density substances, i.e. ESBT-NB and ESBT, which represented skeletal tissue (Figure 2). Where materials had a density which was either identical to or lower than that of water, i.e. the soft tissue equivalents, ESST-NB and ESST, only a modest variation in CT number was noted. In the high-energy image acquisitions, the CT number was more consistent for the different substances than for image acquisitions at lower energies. Thus, the CT numbers documented for the current specimens lay within the CT number spectra for soft tissue and skeletal tissue and concurred with earlier published data [43-45].

The electron densities and average CT numbers obtained at an energy level of 52 keV for the various phantom components are presented in Table 4. Hounsfield units (HU) were used to define the CT numbers; these are affected by material density, and their scale extends from -1000 to +1000, which reflect darker and lighter hues and which, in turn, indicate materials with lower and higher densities, respectively. Air has a HU value of -1000, whereas the HU value for water is zero. The HU value spectra for lower density soft tissue and higher density tissue, e.g. bone, are -700 to +25, and +226 to +3071, respectively [44, 46].

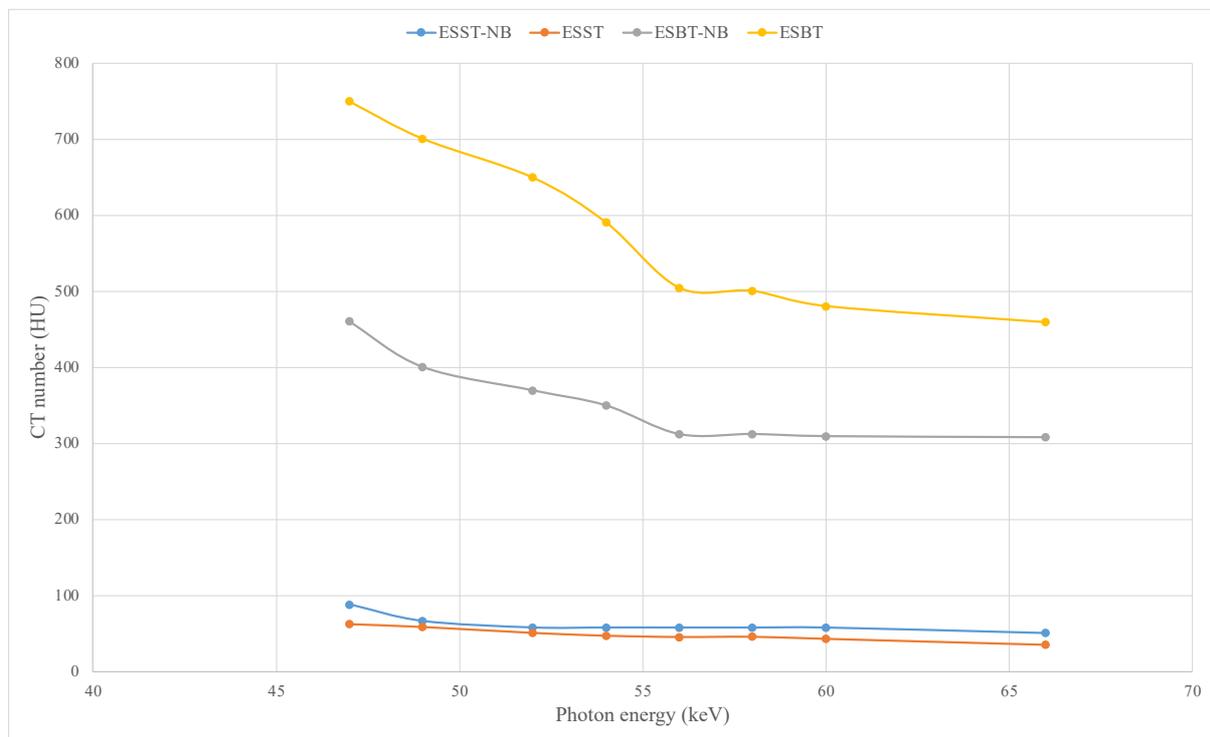

**Figure 2.** CT numbers for different USM tissue-equivalent substitutes as a function of diagnostic photon energy range (47 to 66 keV).

**Table 4.** The mean CT number and the electron density for each plug phantom at 52 keV.

| Plug phantom | Mean CT number | Electron density x $10^{23}$ (electron/cm$^3$) |
|---|---|---|
| Water | 4 | 3.3679 |
| ESST-NB | 58.05 | 3.4431 |
| ESST | 50.996 | 3.4333 |
| Muscle | 69 | 3.4583 |
| Liver | 45 | 3.4249 |
| Bone 200 | 335.42 | 3.8291 |
| ESBT-NB | 370.134 | 3.8774 |
| ESBT | 650 | 4.2668 |
| Bone 800 | 1071 | 4.8526 |
| Bone 1250 | 1651.8 | 5.6608 |

The proximity of the CT numbers for a pair of substances indicated similarity in the attenuation characteristics. Using the mean CT number acquired, the CT numbers giving the best match to soft tissue were documented in relation to the USM ESST-NB and ESST specimens, implying that their attenuation features were in keeping with the CIRS phantom's liver and muscle. The ESBT-NB and ESBT specimens provided the CT numbers which were most tightly in accordance with bone tissue. Equivalent attenuation characteristics were noted between ESBT-NB and ESBT, and bone 200 and bone 800, respectively. The electron density curve was used to acquire and to compute the individual electron densities (Figure 3). An excellent linear fit was demonstrated for CT number (HU) against electron density (linear regression, $R^2 = 1$).

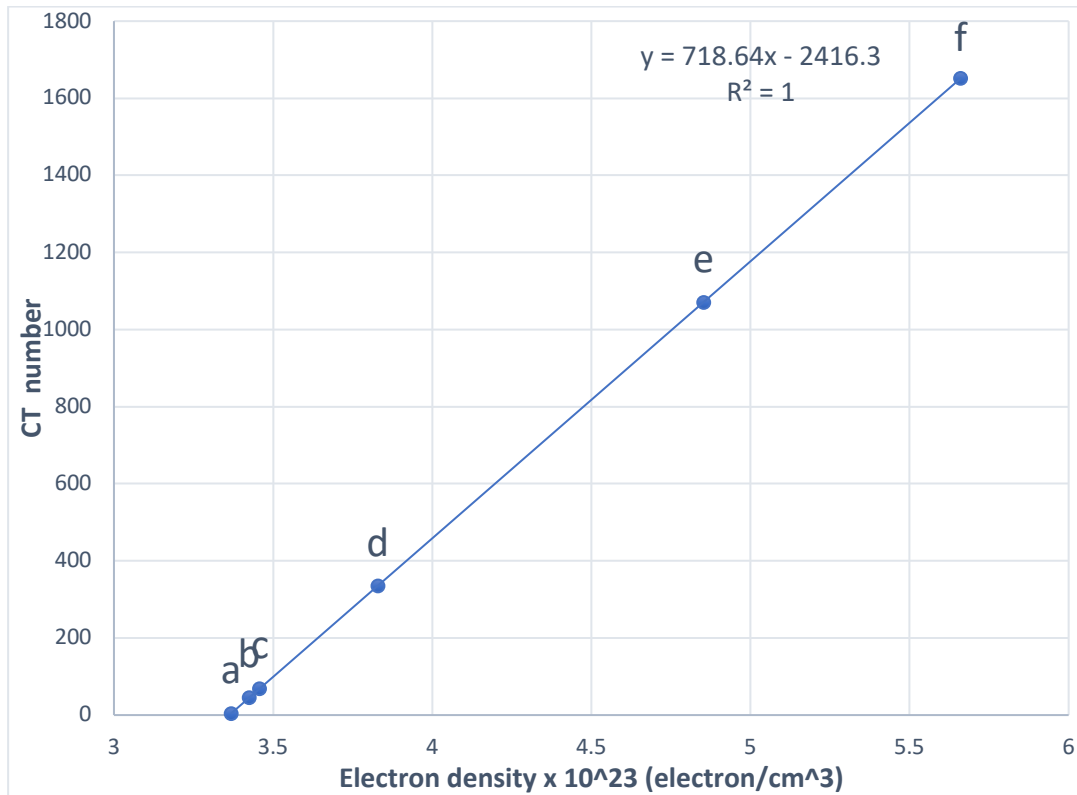

**Figure 3.** Electron density calibration curve against the CT number at tube voltage 52 keV for [a]Water, [b]Muscle, [c] Liver, [d]Bone 200, [e]Bone 800, and [f]Bone 1250.

### *3.3 Mass attenuation coefficient*

The mass attenuation coefficient ratios, $\mu/\rho$ for the different TE materials relating to the newborn and the ORNL reference samples for the targeted photon energy spectrum are compared in Figure 4. Across the energy spectrum, 47 keV to 66 keV, the mass attenuation coefficient ratios for ESST-NB and ESBT-NB show discrepancies with respect to the reference value of -2.03% to -3.01%, and +1.6% to -2.1%, respectively. This arose as priority was given to achieving similar mass densities and mass energy attenuation coefficients for the TE materials in relation to the reference standards.

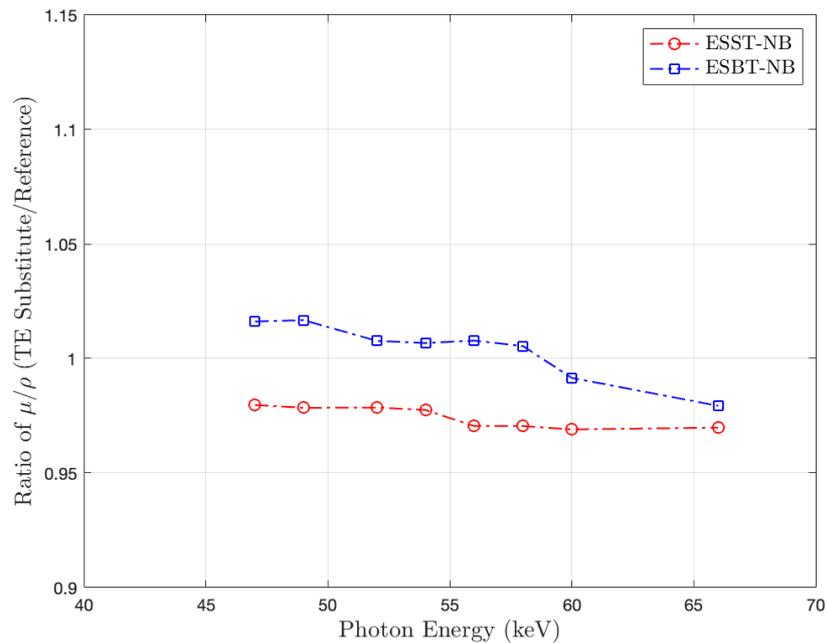

**Figure 4.** The ratio of mass attenuation coefficient μ/ρ to their corresponding reference tissues for the newborn phantom model as a function of diagnostic energy.

Figure 5 illustrates the mass-energy attenuation coefficient ratios, $\mu_{en}/\rho$ for the newborn TE materials. For EEST-NB, values of +1.1% to -1.4% were observed over the spectrum 47 keV - 66 keV, apart from a discrepancy of +1.9% noted at 49 keV in relation to the reference sample values. The equivalent data for ESBT-NB were +1.15% to -0.01% across 47 keV to 66 keV. Within this energy spectrum, the deviation of these values from baseline for the two TE materials representative of a neonate was <±2 in each case.

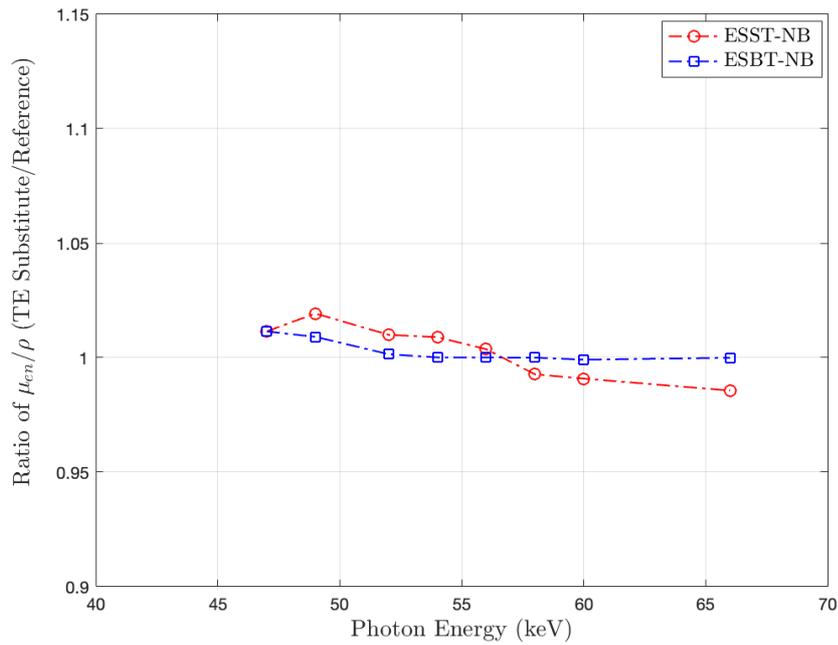

**Figure 5**. The ratio of mass - energy attenuation coefficient $\mu_{en}/\rho$ to their corresponding reference tissues for the newborn phantom model as a function of diagnostic energy

The equivalent comparative data for the TE materials, ESST and ESBT, together with the reference samples from ORNL are presented in Figures 6 (μ/ρ) and 7 (μ$_{en}$/ρ). Across 47 keV to 66 keV, the variations in μ/ρ seen for ESST and ESBT with respect to the reference values are -0.17% to -3.02%, and +1.09% to +0.1%, respectively, and for μ$_{en}$/ρ, +1.92% to -2.53% and +0.3% to -2%, respectively.

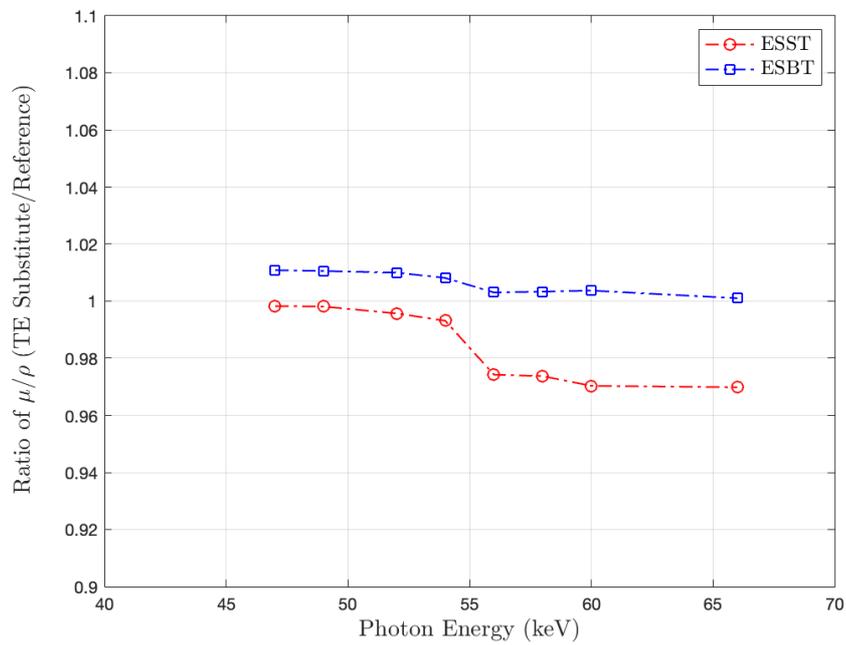

**Figure 6.** The ratio of mass attenuation coefficient $\mu/\rho$ for the ESST and ESBT equivalent tissues to their corresponding reference attenuation values for ORNL (5-year-old model) as a function of diagnostic energy.

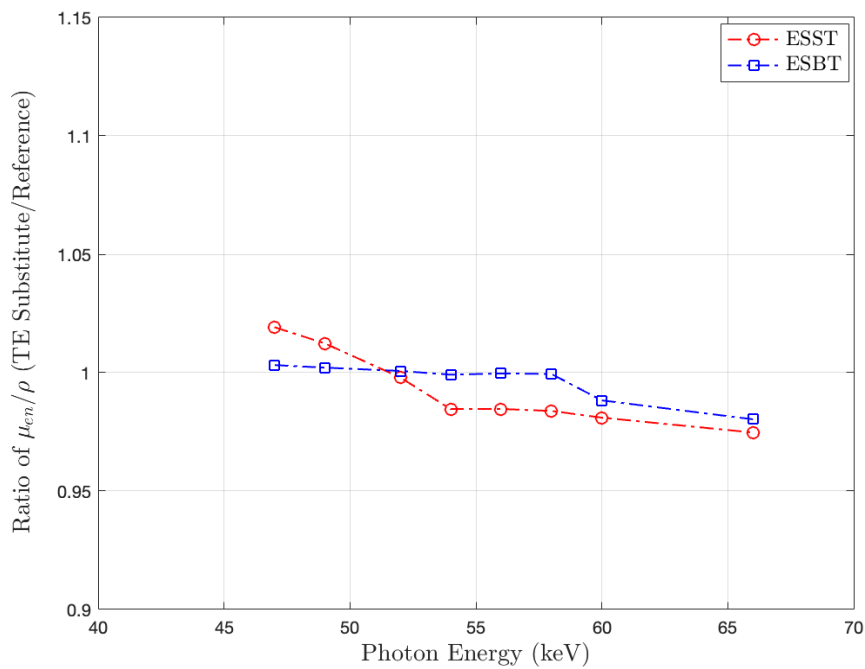

**Figure 7.** The ratio of mass energy attenuation coefficient $\mu_{en}/\rho$ for the ESST and ESBT equivalent tissues to their corresponding reference attenuation values for ORNL (5-year-old model) as a function of diagnostic energy.

### *3.4 Comparison of USM TE substitutes to other reference materials*

Figure 8 provides a comparison of the μ/ρ and μ$_{en}$/ρ ratios for ESST–NB, acrylic and the relevant ORNL reference samples [26] across the desired diagnostic energy range. Since the TE materials innovated by White et al. [11] were not aimed towards representing neonatal subjects, no comparison of ESST–NB with MS11 was performed. With respect to μ/ρ and μ$_{en}$/ρ there is an approximation of acrylic towards the values for ESST–NB within the energy spectra, 47 keV to 66 keV, and > 66 keV, respectively. The equivalent data for ESBT–NB and aluminium are presented in Figure 9, with identical results.

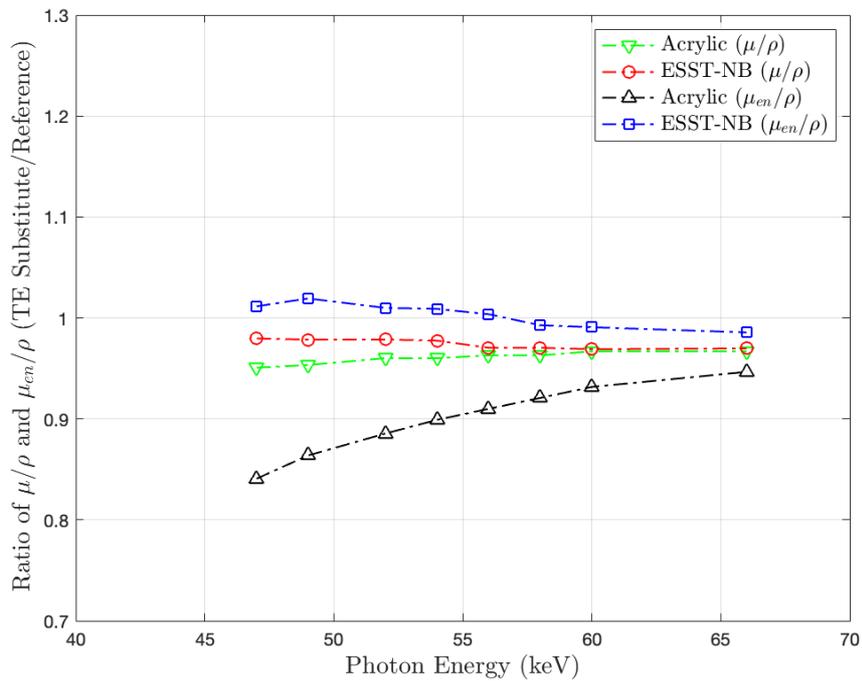

**Figure 8.** Ratios of $\mu/\rho$ and $\mu_{en}/\rho$ for ESST-NB and acrylic to their corresponding reference values ORNL newborn model as a function of diagnostic energy.

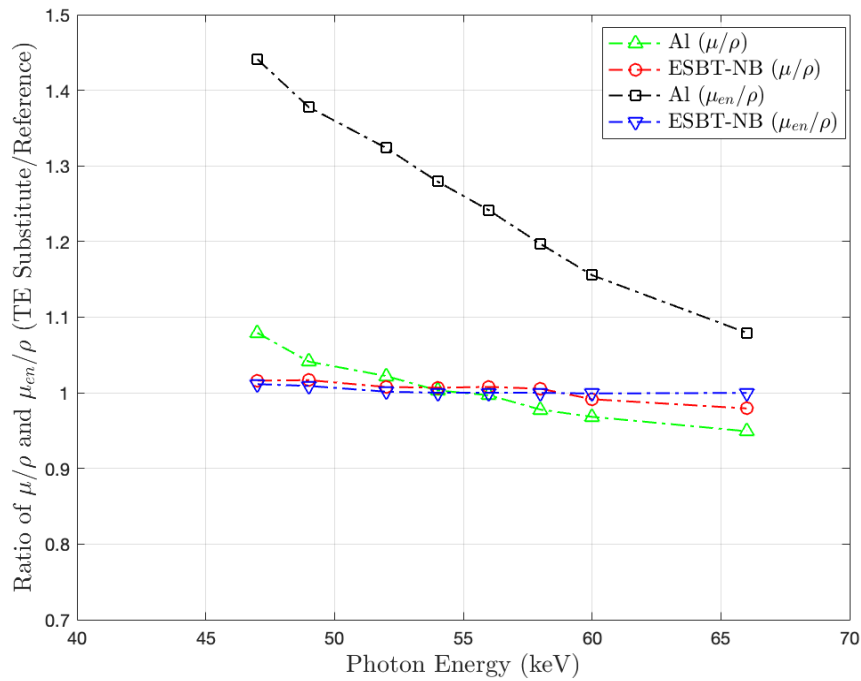

**Figure 9.** Ratios of $\mu/\rho$ and $\mu_{en}/\rho$ for ESBT-NB and aluminum to their corresponding reference values ORNL newborn model as a function of diagnostic energy.

The ESST correspondence with acrylic and MS11, the muscle TE developed by White et al., is shown in Figure 10. Overall, each substance demonstrated concordance with the reference interaction coefficients for soft tissue in a subject aged five years for the required energy spectrum. However, at the lower end of the energy range, μ/ρ and μ$_{en}$/ρ for acrylic were consistently lower in relation to the relevant reference values but demonstrated more concordance at 66 keV. For this subject age range, and for all photon energies tested, the same ratios for ESST and MS11 were approximately equivalent to the ORNL reference samples although for MS11, the agreement for μ$_{en}$/ρ was greater at higher photon energies.

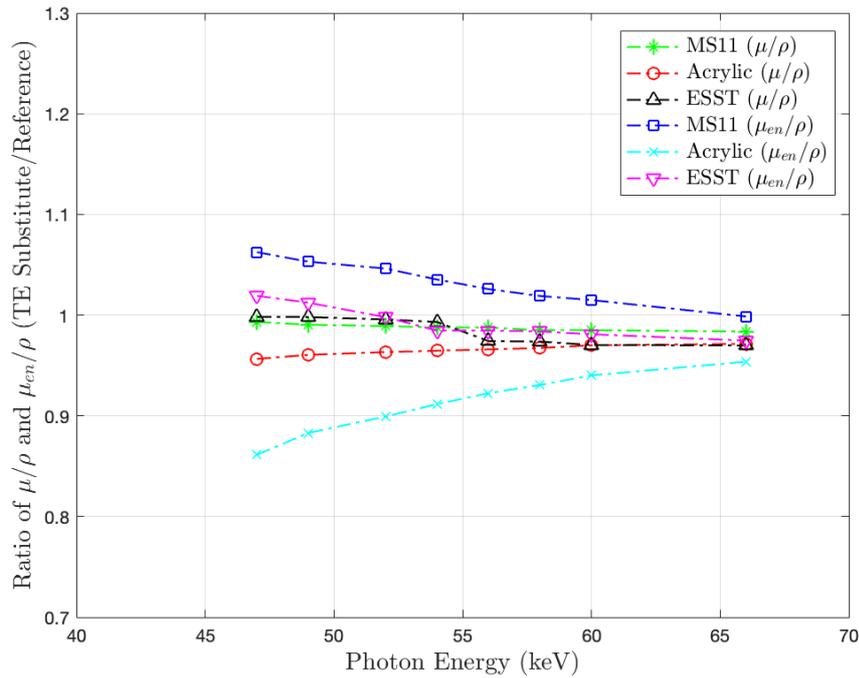

**Figure 10.** Ratios of $\mu/\rho$ and $\mu_{en}/\rho$ for ESST, MS11, and acrylic to their corresponding reference values of ORNL (5-year-old model) as a function of diagnostic energy.

The comparative data for the skeletal TE materials are provided in Figures 11 ($\mu/\rho$) and Figure 12 ($\mu_{en}/\rho$). Additional materials to the ESBT included in this comparison comprise the ORNL reference skeletal sample for a 5-year-old subject, IB1 and IB5, the compounds used by White et al. to represent the spongiosa and cortical bone, respectively, and aluminium. An additional plotted data set is represented, which includes proportional weighting for IB1 and SB5 typical of Reference Man cortical and trabecular bone [47, 48]. Agreement discrepancies were evident in the plots for each material at the lower photon energy range with the exception of ESBT, which demonstrated equivalence for the assessed ratios throughout the photon energy spectrum. Agreement of the $\mu/\rho$ and $\mu_{en}/\rho$ ratios was enhanced with the reference data at higher photon energies. For the energy points assessed, i.e. 47, 49, 52, 54, 56, 58, 60, and 66 keV, ESBT demonstrated close correspondence with the ORNL reference data for bone representative of a subject aged five years.

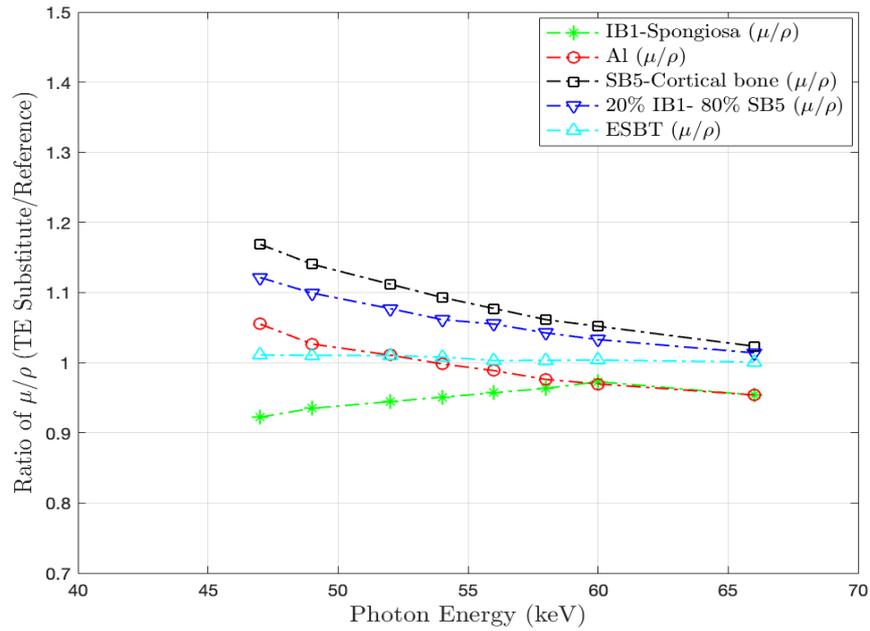

**Figure 11**. Ratios of $\mu/\rho$ for ESBT, IB1, SB5, weighted combination of IB1 and SB5, and aluminum to their corresponding reference values of ORNL (5-year-old) as a function of photon energy.

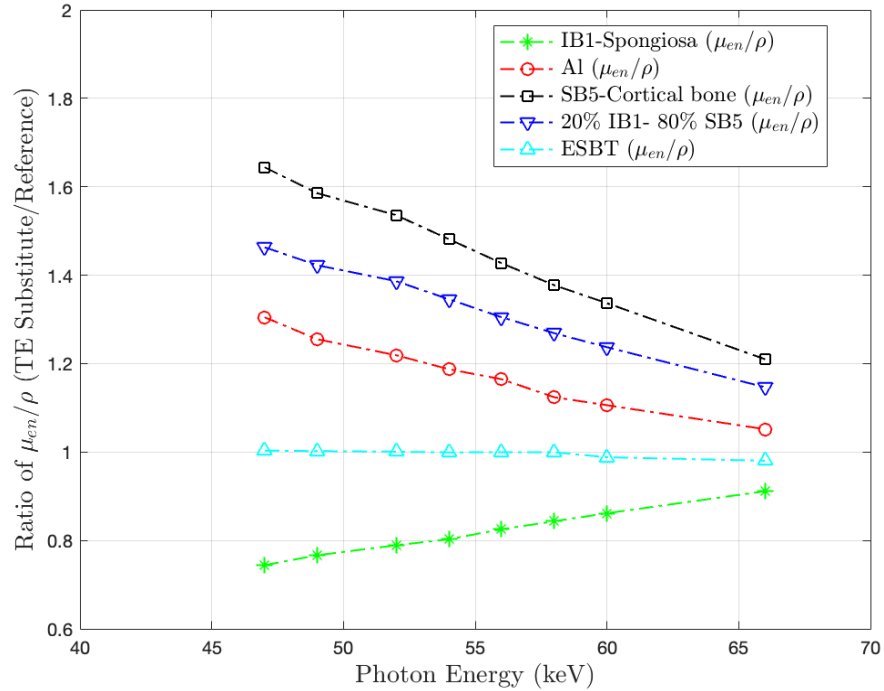

**Figure 12.** Ratios of $\mu_{en}/\rho$ for ESBT, IB1, SB5, weighted combination of IB1 and SB5, and aluminum to their corresponding reference values of ORNL (5-year-old) as a function of photon energy range

## 4 Discussion

A total of four TE materials for use in the creation of phantoms applied for visceral dose approximation during diagnostic CT imaging in paediatric subjects are presented [49, 50]. The properties of the compounds, ESST–NB, ESBT–NB, ESST and ESBT are detailed; these TE materials mimic the radiographic characteristics of soft tissue and skeletal tissue in a neonate and child aged five years. The elemental proportions, published by Cristy and Eckerman [26] in relation to the ORNL newborn stylized computational model, were utilised as a benchmark during the manufacturing process.

The $Z_{eff}$ and $Z_{eff}^{PE}$ values for ESST-NB, ESBT-NB, ESST, and ESBT show discrepancies with respect to the reference values of $\leq 3$, as shown in Tables 2 and 3. The closest correlations of the USM TE material CT numbers were demonstrated between ESST-NB and the CIRS liver phantom, ESST and the CIRS muscle phantom, ESBT-NB and bone 200, and ESBT and bone 800 (Figure 2). Additionally, as shown in Table 4, the electron density values were concordant with the CIRS phantom electron density data.

For the $\mu/\rho$ ratio, when compared to the reference data, discrepancies of -2.03% to -3.0% and +1.6% to -2.1% were noted for ESST-NB and ESBT-NB, respectively, over the tested energy spectrum of 47 keV to 66 keV (Figure 4). These TE materials showed deviations from the reference values in terms of $\mu_{en}/\rho$ of < ±2 for the same energy values (Figure 5). The comparative data for these ratios in relation to ESST and ESBT are illustrated in Figures 6 and 7. Over the above energy spectrum, the greatest discrepancies for $\mu/\rho$ were +1.6% to –3.01% and for $\mu_{en}/\rho$ +1.15% to –1.4%. For the TE materials representative of a 5-year-old, the respective equivalent data ranges were +1.09 % to – 3.02% and +1.92% to –2.53%.

Of note, was that the attenuation data presented here exhibited lower values than those published by Jones et al. [13] over a similar energy spectrum. This is likely to reflect the use of Araldite GY-250 in this study.

## 5    Conclusion

The de novo TE materials, ESST-NB, ESBT-NB, ESST and ESBT, designed to represent newborn and 5-year-old paediatric soft tissue and skeletal tissue, and produced using a technique based on epoxy resin, were shown to be suitable for purpose. They are therefore promising compounds for use in physical phantom generation for diagnostic radiology applications and for quality control checks in clinical X-ray apparatus within the lower energy spectrum, i.e. 47 keV to 66 keV. They offer an option to the regular ACR phantom applied and have potential advantages for dose minimisation.

**Acknowledgment**

The authors acknowledged the Dr. Ranjeet Bhagwan Singh Medical Research Grant 2019/2020 (awarded to Dr. Nurul Hashikin Ab. Aziz) for the financial support in conducting this research.

**Competing interests**

None declared

**Funding**

This work was funded by Dr. Ranjeet Bhagwan Singh Medical Research Grant 2019/2020 (awarded to Dr. Nurul Hashikin Ab. Aziz).

**Ethical approval**

Not required

# References

1. Council, N.R., *Health risks from exposure to low levels of ionizing radiation: BEIR VII phase 2.* 2006.
2. Brenner, D.J., et al., *Estimated risks of radiation-induced fatal cancer from pediatric CT.* American journal of roentgenology, 2001. **176**(2): p. 289-296.
3. Jones, A., et al., *Tomographic physical phantom of the newborn child with real-time dosimetry I. Methods and techniques for construction.* Medical physics, 2006. **33**(9): p. 3274-3282.
4. Murphy, P.H., *NCRP Report No. 100, Exposure Of The Us. Population From Diagnostic Medical Radiation: Maryland, National Council on Radiation Protection & Measurements. 1989, 103 pp. $14.00.* 1989, Soc Nuclear Med.
5. Bower, M.W., *A physical anthropomorphic phantom of a one year old child with real-time dosimetry.* 1997: University of Florida.
6. Kienbock, R., *On the quantimetric method.* Archives of the Roentgen Ray, 1906. **11**(1): p. 17-20.
7. White, D., *The formulation of substitute materials with predetermined characteristics of radiation absorption and scattering.* 1974, Queen Mary University of London.
8. Constantinou, C., *Tissue substitutes for particulate radiations and their use in radiation dosimetry and radiotherapy.* 1978.
9. Constantinou, C., F. Attix, and B.R. Paliwal, *A solid water phantom material for radiotherapy x-ray and γ-ray beam calibrations.* Medical physics, 1982. **9**(3): p. 436-441.
10. White, D., *The formulation of tissue substitute materials using basic interaction data.* Physics in Medicine & Biology, 1977. **22**(5): p. 889.
11. White, D., R. Martin, and R. Darlison, *Epoxy resin based tissue substitutes.* The British journal of radiology, 1977. **50**(599): p. 814-821.
12. White, D., *Tissue substitutes in experimental radiation physics.* Medical physics, 1978. **5**(6): p. 467-479.
13. Jones, A., D. Hintenlang, and W. Bolch, *Tissue-equivalent materials for construction of tomographic dosimetry phantoms in pediatric radiology.* Medical physics, 2003. **30**(8): p. 2072-2081.
14. Winslow, J.F., et al., *Construction of anthropomorphic phantoms for use in dosimetry studies.* Journal of Applied Clinical Medical Physics, 2009. **10**(3): p. 195-204.
15. Akhlaghi, P., H.M. Hakimabad, and L.R. Motavalli, *Determination of tissue equivalent materials of a physical 8-year-old phantom for use in computed tomography.* Radiation Physics and Chemistry, 2015. **112**: p. 169-176.
16. Badiuk, S.R., D.K. Sasaki, and D.W. Rickey, *An anthropomorphic maxillofacial phantom using 3-dimensional printing, polyurethane rubber and epoxy resin for dental imaging and dosimetry.* Dentomaxillofacial Radiology, 2022. **51**(1): p. 20200323.

17. Yücel, H. and A. Safi, *Investigation of the suitability of new developed epoxy based-phantom for child's tissue equivalency in paediatric radiology.* Nuclear Engineering and Technology, 2021. **53**(12): p. 4158-4165.
18. Hermann, K.-P., et al., *Polyethylene-based water-equivalent phantom material for x-ray dosimetry at tube voltages from 10 to 100 kV.* Physics in Medicine & Biology, 1985. **30**(11): p. 1195.
19. Homolka, P. and R. Nowotny, *Production of phantom materials using polymer powder sintering under vacuum.* Physics in Medicine & Biology, 2002. **47**(3): p. N47.
20. Suess, C., W.A. Kalender, and J.M. Coman, *New low-contrast resolution phantoms for computed tomography.* Medical physics, 1999. **26**(2): p. 296-302.
21. Iwashita, Y., *Basic study of the measurement of bone mineral content of cortical and cancellous bone of the mandible by computed tomography.* Dentomaxillofacial Radiology, 2000. **29**(4): p. 209-215.
22. *Huntsman Corporation is an American multinational manufacturer and marketer of chemical products for consumers and industrial customers*. Accessed December 2022; Available from: https://www.huntsman.com/products.
23. *True Phantom Solutions, Pediatric Full Human Body Phantom for X-Ray CT & MRI Training.* Accessed December 2022; Available from: https://truephantom.com/product/pediatric-full-human-body-phantom-for-x-ray-ct-mri-training/.
24. *Kyoto Kagaku (2022) Newborn WholeBody Phantom "PBU-80"*. Accessed December 2022; Available from: https://www.kyotokagaku.com/en/products_data/ph-50b/.
25. *Computerized Imaging Reference Systems (CIRS) Inc (2022). Pediatric Anthropomorphic Training Phantoms, Model 715 series.* Accessed December 2022; Available from: https://www.cirsinc.com/products/x-ray-fluoro/pediatricanthropomorphic-training-phantoms/.
26. Cristy, M.E., K. F., *Specific absorbed fractions of energy at various ages from internal photon sources.* 1987: United States.
27. Bolch, W., et al., *ICRP Publication 143: paediatric reference computational phantoms.* Annals of the ICRP, 2020. **49**(1): p. 5-297.
28. Attix, F.H., *Introduction to radiological physics and radiation dosimetry*. 2008: John Wiley & Sons.
29. Johns, H.E. and J.R. Cunningham, *The physics of radiology.* 1983.
30. Siegelman, S.S., et al., *CT of the solitary pulmonary nodule*, in *Radiology Today 1*. 1981, Springer. p. 113-120.
31. Maia, R.S., et al., *An algorithm for noise correction of dual-energy computed tomography material density images.* International journal of computer assisted radiology and surgery, 2015. **10**(1): p. 87-100.
32. Hubbell, J., *Tables of X-ray mass attenuation coefficients 1 keV to 20 MeV for elements Z= 1 to 92 and 48 additional Substance of dosimetric Interest.* NISTIR 5632, 1995.
33. Seltzer, S.M., *Calculation of photon mass energy-transfer and mass energy-absorption coefficients.* Radiation research, 1993. **136**(2): p. 147-170.
34. Dick, C., C. Soares, and J. Motz, *X-ray scatter data for diagnostic radiology.* Physics in Medicine & Biology, 1978. **23**(6): p. 1076.
35. Motz, J. and C. Dick, *X-ray scatter background signals in transmission radiography.* Medical Physics, 1975. **2**(5): p. 259-267.

36. Fetterly, K.A. and N.J. Hangiandreou, *Effects of x-ray spectra on the DQE of a computed radiography system.* Medical physics, 2001. **28**(2): p. 241-249.
37. Conway, B., et al., *A patient-equivalent attenuation phantom for estimating patient exposures from automatic exposure controlled x-ray examinations of the abdomen and lumbo–sacral spine.* Medical physics, 1990. **17**(3): p. 448-453.
38. Mah, E., E. Samei, and D.J. Peck, *Evaluation of a quality control phantom for digital chest radiography.* Journal of applied clinical medical physics, 2001. **2**(2): p. 90-101.
39. Chotas, H.G., et al., *Quality control phantom for digital chest radiography.* Radiology, 1997. **202**(1): p. 111-116.
40. Chotas, H.G., et al., *Small object contrast in AMBER and conventional chest radiography.* Radiology, 1991. **180**(3): p. 853-859.
41. Scheck, R., et al., *Radiation dose and image quality in spiral computed tomography: multicentre evaluation at six institutions.* The British Journal of Radiology, 1998. **71**(847): p. 734-744.
42. McCollough, C.H. and F.E. Zink, *Performance evaluation of a multi-slice CT system.* Medical physics, 1999. **26**(11): p. 2223-2230.
43. Watanabe, Y., *Derivation of linear attenuation coefficients from CT numbers for low-energy photons.* Physics in Medicine & Biology, 1999. **44**(9): p. 2201.
44. Chougule, V., A. Mulay, and B. Ahuja, *Clinical case study: spine modeling for minimum invasive spine surgeries (MISS) using rapid prototyping.* Bone (CT), 2018. **226**: p. 3071.
45. Yagi, M., et al., *Gemstone spectral imaging: determination of CT to ED conversion curves for radiotherapy treatment planning.* Journal of applied clinical medical physics, 2013. **14**(5): p. 173-186.
46. Sichel, J.-Y., et al., *Artifactual thickening of the sinus walls on computed tomography: A phantom model and clinical study.* Annals of Otology, Rhinology & Laryngology, 2000. **109**(9): p. 859-862.
47. 2, I.C.o.R.P.C., *Limits for Intakes of Radionuclides by Workers: A Report*. Vol. 2. 1979: International Commission on Radiological Protection.
48. Icrp, *Report of the Task Group on Reference Man: A Report*. 1975, Pergamon Oxford.
49. Nipper, J., J. Williams, and W. Bolch, *Creation of two tomographic voxel models of paediatric patients in the first year of life.* Physics in Medicine & Biology, 2002. **47**(17): p. 3143.
50. Sessions, J., et al., *Comparisons of point and average organ dose within an anthropomorphic physical phantom and a computational model of the newborn patient.* Medical Physics, 2002. **29**(6): p. 1080-1089.